# Binomial and ratio-of-Poisson-means frequentist confidence intervals applied to the error evaluation of cut efficiencies

Gioacchino Ranucci
*Istituto Nazionale di Fisica Nucleare*
*Via Celoria 16 - 20133 Milano*
*e-mail: gioacchino.ranucci@mi.infn.it*

## Abstract

The evaluation of the error to be attributed to cut efficiencies is a common question in the practice of experimental particle physics. Specifically, the need to evaluate the efficiency of the cuts for background removal, when they are tested in a signal-free-background-only energy window, originates a statistical problem which finds its natural framework in the ample family of solutions for two classical, and closely related, questions, i.e. the determination of confidence intervals for the parameter of a binomial proportion and for the ratio of Poisson means. In this paper the problem is first addressed from the traditional perspective, and afterwards naturally evolved towards the introduction of non standard confidence intervals both for the binomial and Poisson cases; in particular, special emphasis is given to the intervals obtained through the application of the likelihood ratio ordering to the traditional Neyman prescription for the confidence limits determination. Due to their attractiveness in term of reduced length and of coverage properties, the new intervals are well suited as interesting alternative to the standard Clopper-Pearson PDG intervals.

## 1. Introduction

The problem tackled in the present work is that typically encountered in determinations of cut efficiencies for background rejections while searching for a signal. In investigations of this kind the experimenter, after performing the estimate of the efficiency, faces the problem to attach to the number, comprised between 0 and 1, just obtained the proper uncertainty, usually defined as the corresponding 68.27% range in the context of the Neyman construction of confidence intervals (i.e. 1 $\sigma$ interval for analogy to the Gaussian case). For example, if a researcher is looking for a signal in an observation window affected by noise, common practice is that he/she designs some cuts to suppress the background whilst preserving the signal itself. Obviously he/she wants to know how efficient is his/her procedure in removing the unwanted counts. For this purpose he/she may select a window in an adjacent energy region of the same width, in which only background is expected, and calibrate there the efficiency of the cuts against noise. So he/she evaluates a quantity $p$ which is the ratio of the background surviving the cuts over the whole background counts. Then he/she wants to define the uncertainty to his/her result, the only a-priori information available being the average number of background counts in the observation window (this is assumed to be equal both in the observation and test windows). Thus, in summary, the problem is to set a confidence interval on the fraction $p$ of success of $k$ surviving counts out of total $n$, being $n$ Poisson fluctuating with known mean value $\mu$ and $k$ Poisson fluctuating with unknown mean $p\mu$.

With the above formulation, it is easy to recognize the similarity of this problem to that regarding the estimation of the confidence intervals of the parameter $p$ of a binomial proportion, the major difference being that in the binomial case the total number of events $n$ is fixed.

Therefore, the first part of the paper is devoted to gain some new insight in the construction of the confidence intervals in the framework of the binomial problem; in doing so not only the standard Clopper-Pearson [1] solution reported by the PDG [2] is outlined, but more significantly two non standard solutions are described, based, respectively, on the Crow-Gardner [3] and the likelihood ratio ordering [4] methods, with particular focus on the advantages that they ensure with respect to the Clopper-Pearson intervals.

Afterwards, the case in which the total number of events $n$ fluctuates is introduced naturally, as a generalization of the binomial problem; in a first instance the properties of the previous binomial intervals are reviewed for fluctuating $n$, and finally a full generalization based on the likelihood ratio ordering method is thoroughly described in the case in which $\mu$ is known, originating confidence intervals with extremely good coverage behavior.

It should be emphasized that, from a foundational point of view, this simple problem is extremely interesting, since it originates a perfect and beautiful example of the full frequentist tridimensional construction of confidence intervals via the concept of the likelihood ratio ordering.

Throughout this work the sample population is denoted with $n$, and the number of events featuring the specific attribute under consideration (i.e. survival after the cuts) with $k$. Depending upon the context, $n$ and $k$ may denote either the corresponding random variables or the specific outcomes after a trial. In order to avoid unnecessary complications, the notation has not been differentiated, nevertheless from the flow of the text it should be clear to the reader which is their meaning in each instance in which they are used.

## 2. Confidence limits for *n* fixed

Addressing initially the binomial problem, in this paragraph the relevant Clopper-Pearson standard solution is first reviewed, and afterward the two anticipated alternative non standard solutions are presented, as well.

*2a Clopper-Pearson*

It is well known that the binomial problem traces back to the 30's. The classical Clopper-Pearson limits [1] have been introduced in the high energy physics field by James and Roos [5], and are the limits quoted by the PDG [2].

Clopper and Pearson developed their intervals adhering strictly to the original Neyman prescription to construct confidence intervals [6] (see also the methodological explanations reported in [2] and [4]). Such intervals are thus obtained through the construction of the relevant confidence belts, which in turns stem from the identification, for each possible value of the parameter $p$, of the corresponding acceptance regions. Numerically, the limits $k_\beta^{down}$ and $k_\beta^{up}$ of the acceptance region for each $p$ are obtained by solving the following conditions:

$$\sum_{k=0}^{k_\beta^{down}} \binom{n}{k} p^k (1-p)^{n-k} \le \beta \tag{1}$$

$$\sum_{k=k_\beta^{up}}^{n} \binom{n}{k} p^k (1-p)^{n-k} \le \beta \tag{2}$$

where $\beta=(1-\alpha)/2$, and $\alpha$ is defined such that $100\alpha\%$ is the desired confidence level. The Clopper-Pearson limits of PDG, being defined according to equations (1) and (2), are central intervals. As concrete examples, the confidence belts for 10 and 100 trials are shown in Fig. 1 and 2, computed both as 95% and 68.27% central confidence intervals, exploiting numerically the Neyman construction. It is worth to remind the interpretation of these plots: upon performing an experiment in which 10 (100) events are obtained in total, the number of events which survive the specified cuts (or more generally the number of events which fulfill a certain property) are counted. The inference of the intrinsic fraction of events $p$ which in the examined population feature the property under consideration is clearly given by the ratio of the detected events with that quality to the total events; the uncertainty on this evaluation is given by the interval obtained by intersecting the contours in the figures with the vertical segment drawn from the abscissa value corresponding to the number of events with that quality. By construction, the segment drawn from the measured success number and intersecting the confidence belts covers with the stated confidence level the true, unknown value of the success parameter *p*.

Though the confidence belts in Fig. 1 and 2 have been obtained via the numerical calculation which originates from the Neyman construction, it has, however, been checked that they coincide with the analytical formulation reported in the PDG, and which stems from the relation between the beta and F functions and the cumulative binomial distribution.

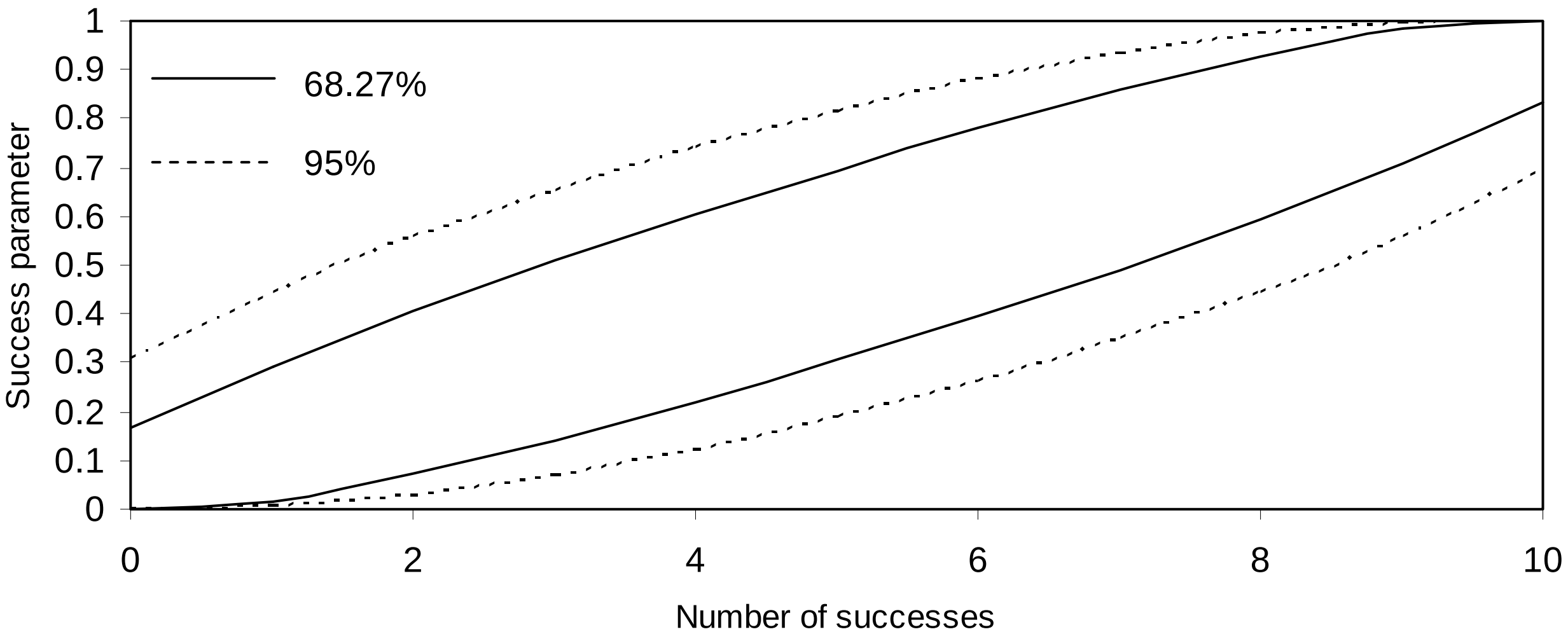

*Fig. 1 – Standard Clopper-Pearson 68.27% and 95% confidence belts for 10 trials*

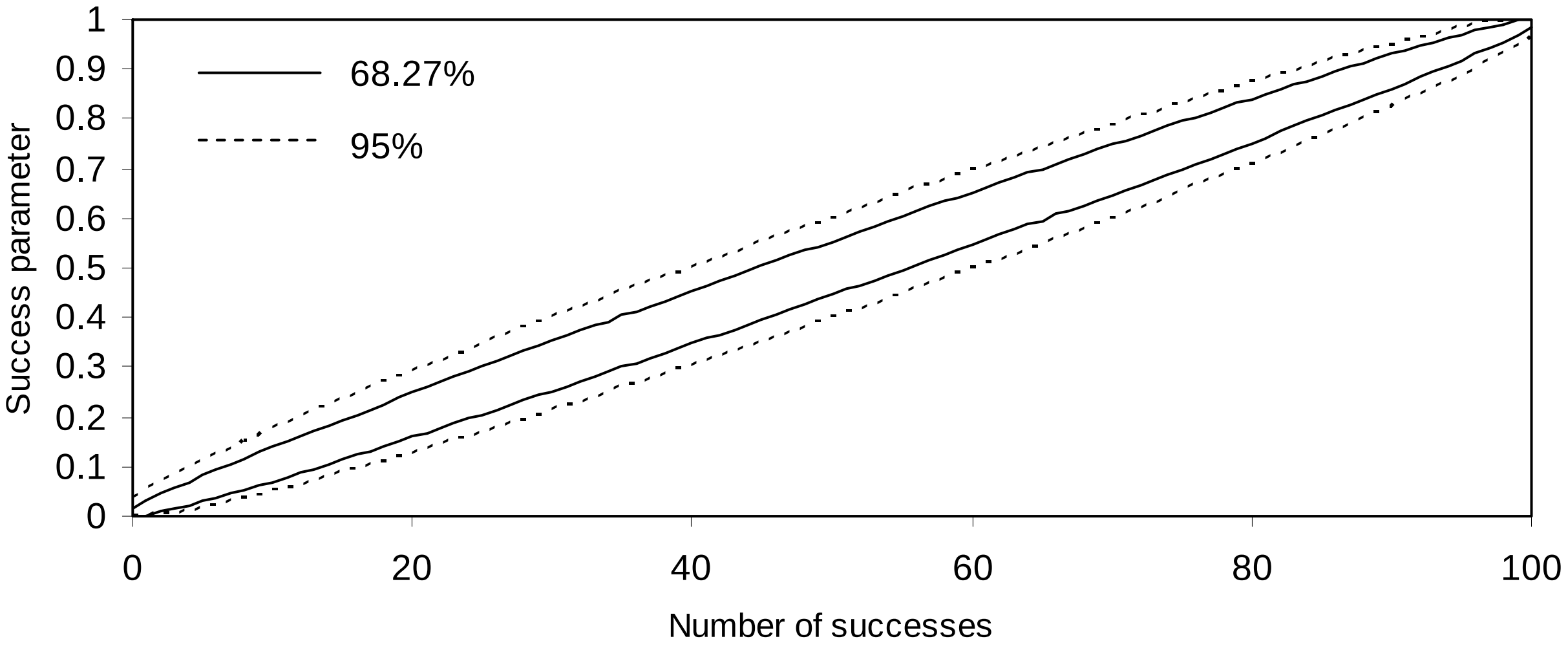

*Fig. 2 – Standard Clopper-Pearson 68.27% and 95% confidence belts for 100 trials*

A well known property of the Clopper-Pearson intervals is that they ensure that for no value of the true parameter *p* there is under-coverage. We remind that under-coverage occurs when for some combination of *p* and *n* the confidence intervals cover the true parameter value less than the specified confidence level. The absence of under-coverage can be appreciated in Fig. 3, where the coverage is reported as function of *p* in the case of 10 trials: the coverage never drops below the nominal value (which in the example are respectively 68.27% and 95%). On the contrary, there may be strong over-coverage, especially for small number of trials, like the case of 10 trials considered here; this is an unfortunate circumstance, unavoidably linked with the discreteness of the binomial distribution. Since in many branches of statistics severe over-coverage is an undesirable property, several alternative confidence intervals have been developed to mitigate this effect, at the cost of some under-coverage for some combinations of *p* and *n*. However, such approximated binomial intervals have not become of common practice in our area because of their under-coverage, and thus they are not discussed here. For recent reviews and works related to them the reader can refer to [7], [8] and [9].

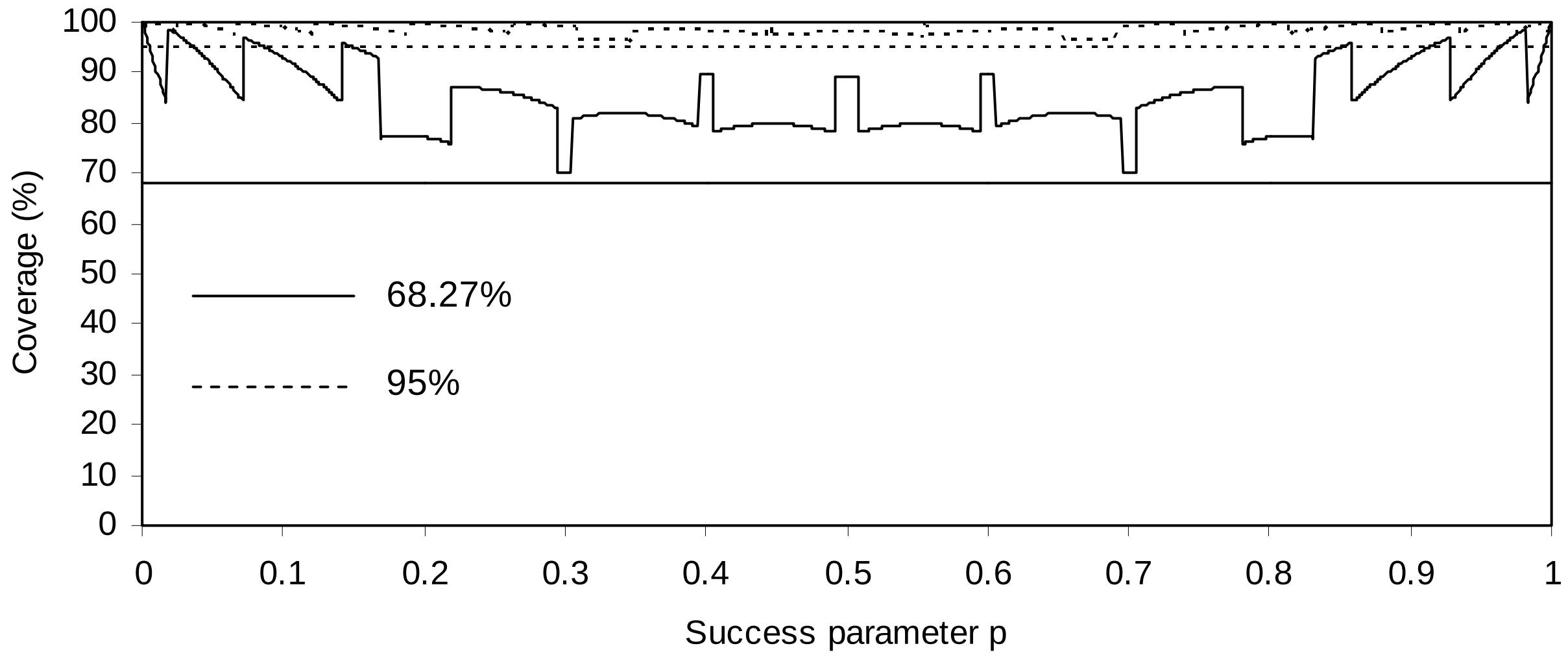

*Fig. 3 –Clopper-Pearson 68.27% and 95% coverage for 10 trials*

Instead, confidence intervals which, as the Clopper-Pearson intervals, do not undercover can be constructed following two alternative approaches discussed by Feldman and Cousins in [4]. These two methods differ with respect to the Clopper-Pearson implementation because of the different criteria adopted to construct, according to the general Neyman prescription, the confidence belt, in particular releasing the constrains that the intervals should be central intervals; these two methods based, respectively, on the Crow-Gardner and on the likelihood ratio ordering principles are illustrated in the next sub-paragraphs.

*2.b Crow Gardner and likelihood ratio confidence limits*

In the Crow-Gardner approach [3] the confidence limits are obtained through acceptance regions which are constructed by adding points in descending order of probability, until the required coverage is met or exceeded.

Instead, according to the Feldman and Cousins likelihood ratio ordering criterion [4], points are added to the acceptance region in descending order of the ratio between the a-priori probability of a point and the likelihood for that point obtained replacing in the probability expression the value of the parameter with the corresponding best fit value.

The Crow Gardner and likelihood ratio confidence belts for 10 trials are shown in Fig. 4, overlapped for comparison to the standard Clopper-Pearson confidence belt; they are computed for 68.27% confidence intervals. The corresponding coverage is plotted in Fig. 5 and 6. Specifically, Fig. 5 shows together the coverage of the Clopper-Pearson and of the Crow-Gardner methods, while Fig. 6 displays the coverage of the Clopper-Pearson and of the likelihood ratio method.

By inspecting Fig. 4, it is interesting to note that, while the general trend of the confidence belt does not change much across the three methods, there are however some appreciable differences. In particular, in the Crow-Gardner case the confidence belt is entirely contained within the Clopper-Pearson belt, thus originating confidence limits always of shorter length, with the exception of the two limiting situation of 0 and 10, for which there is perfect coincidence between the two criteria.

On the other hand, the likelihood ratio belt is almost always contained within the Clopper-Pearson belt, with the two exceptions for 3 and 7 successes. In all other cases, but these two, the likelihood approach, hence, produces confidence intervals shorter than the Clopper-Pearson ones. A very interesting result, deserving to be outlined, is that at the two limiting condition of 0 successes or 10 successes the likelihood method produces substantially tight intervals than the other two methodologies.

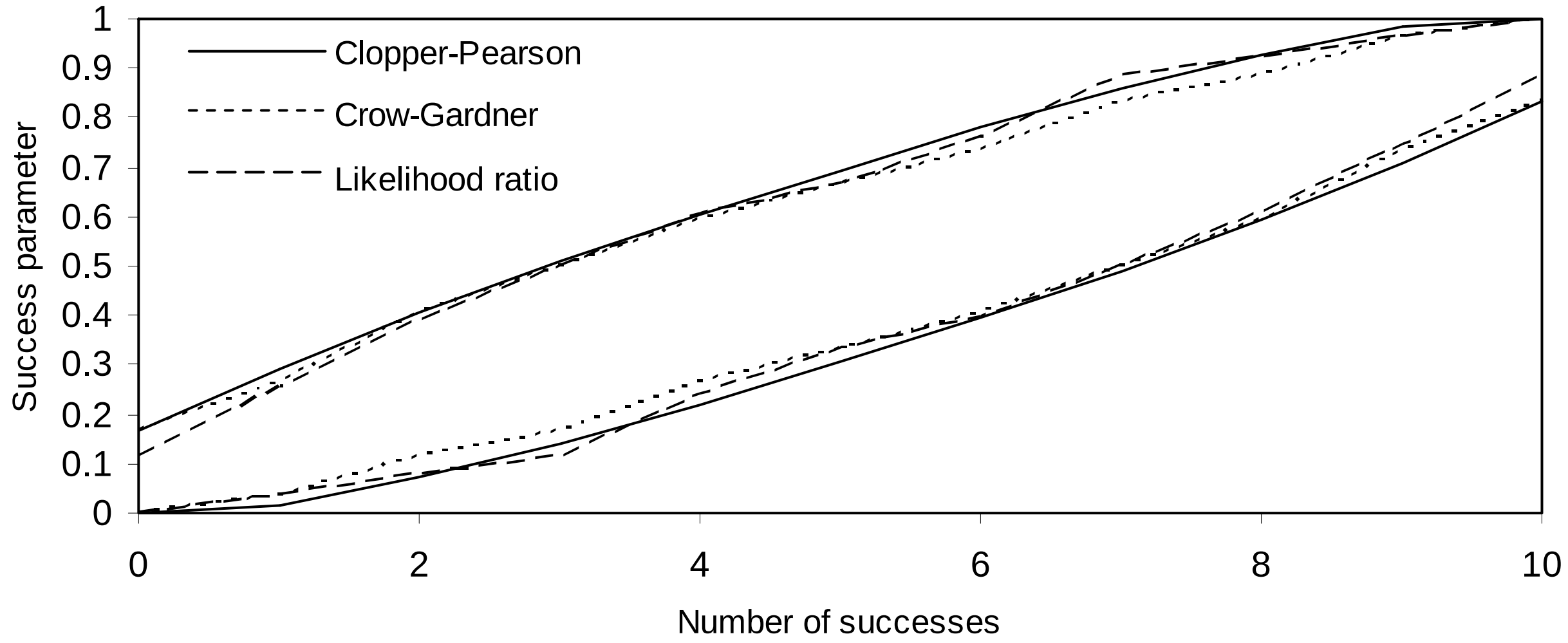

*Fig. 4 – Crow-Gardner and Likelihood ratio 68.27% confidence belts for 10 trials compared with the standard Clopper-Pearson confidence belt*

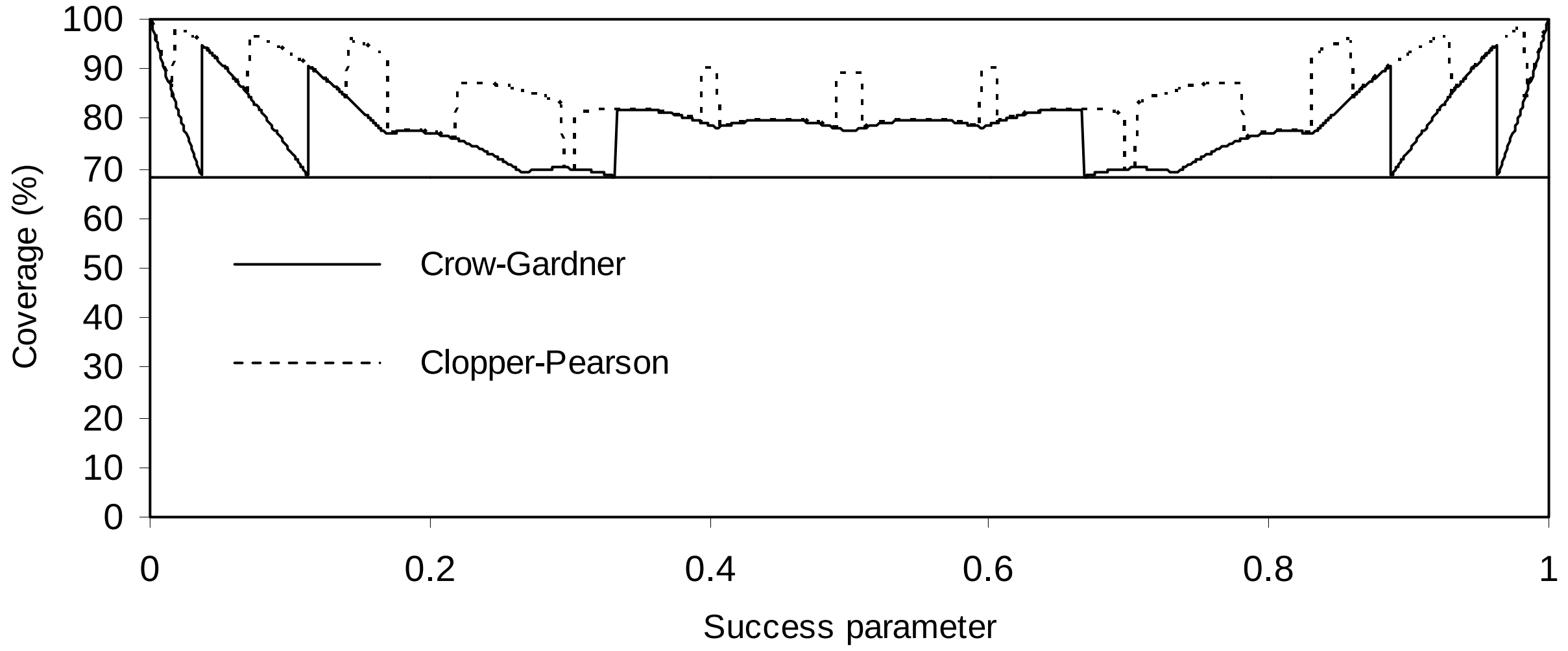

*Fig. 5 – Crow-Gardner and Clopper-Pearson 68.27% coverage for 10 trials*

Coming back to the coverage issue, the two figures 5 and 6 demonstrate that both the Crow-Gardner and the likelihood methods perform better than the standard method, in the sense that they mitigate for several values of the success parameter $p$ the over-coverage of the Clopper-Pearson intervals, still ensuring no under-coverage for any $p$.

*2.c Comparisons with the Wald approximation*

In Fig. 7 there is the comparison, in the case of 10 trials, between the length of the 68.27% intervals obtained for the three methods under consideration, together with the corresponding length of the approximate, and widely used, Wald intervals [10], expressed by

$$2\sqrt{\frac{k}{n^2}\left(1-\frac{k}{n}\right)} \quad (3).$$

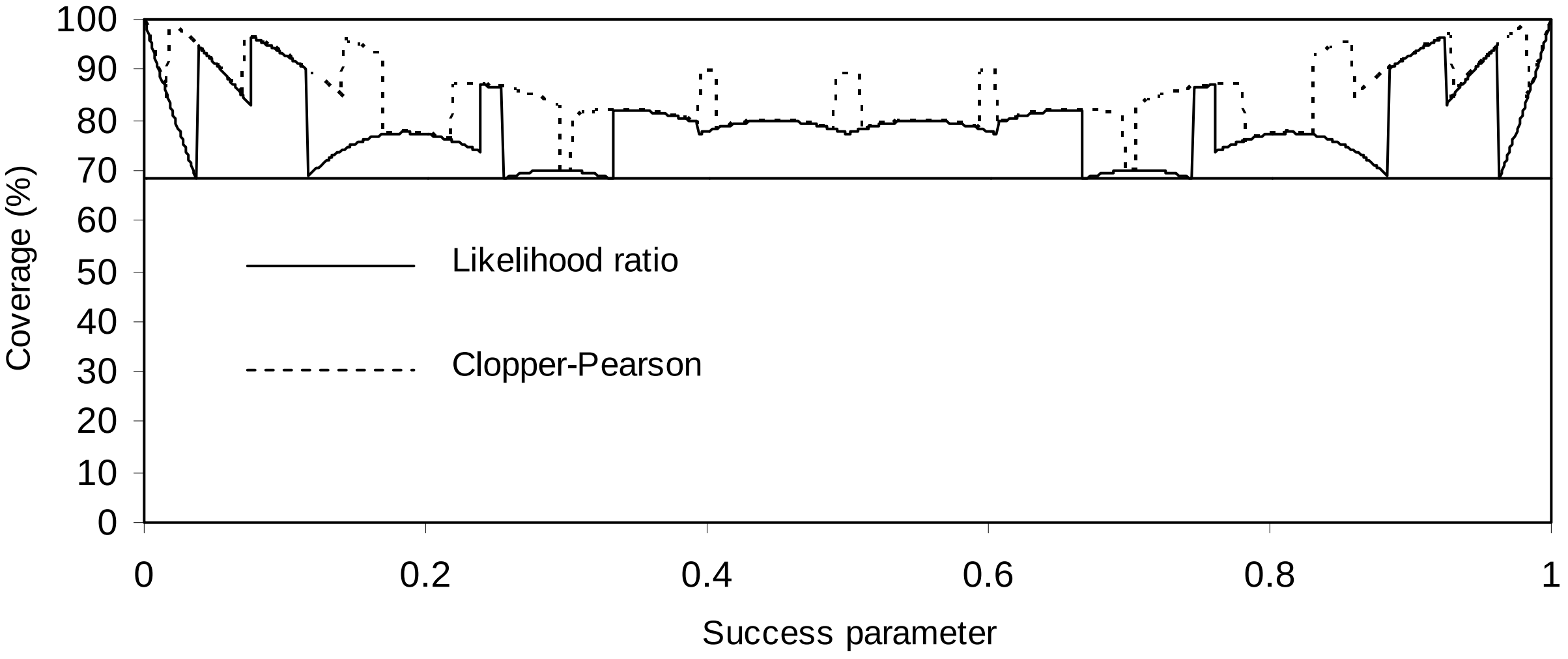

*Fig. 6 – Likelihood ratio and Clopper-Pearson 68.27% coverage for 10 trials*

Clearly, as it is well known, this approximation fails at the boundaries for $k$=0 and $k$=$n$. The comparison among the three methods in Fig. 7 elucidates better the conclusions already reported in the previous subparagraph, i.e. that the Crow Gardner method performs better than the Clopper-Pearson limits with the exception of the extreme cases, and that the likelihood ratio approach is almost always better than the Clopper-Pearson, with the remarkable tightening of the intervals at the extremes. The comparison with the Wald intervals shows, instead, that the they can be regarded only as a lower limit of the true intervals. The study of the coverage of the Wald approximation for the binomial problem is not performed here (a recent theoretical derivation of the coverage can be found in [11]), but it is well known in the literature [8] that it is affected by severe under-coverage properties which make it not perfectly suited for the binomial problem, especially for low $n$

In Fig. 8, for the sake of completeness, the same comparison is reported for the much more favorable case of 100 trials. Now both the Crow-Gardner and the likelihood ratio methods produce uniformly better and tighter intervals with respect to the standard Clopper-Pearson result. It is to be outlined that at the limiting situations $k$=0 and $k$=n the same circumstance already highlighted in the 10 trial plot occurs, i.e. that the Crow-Gardner and Clopper-Pearson intervals are equal, while the likelihood interval is shorter. Also the approximate Wald formula obviously performs better, especially if compared with the Crow-Gardner and likelihood ratio methods.

### 3. Confidence limits and coverage for fluctuating *n*

Strictly speaking, the previous results are valid when, while repeating the experiment, the same total number of events is obtained. What happens if, on the contrary, the total number of events fluctuate according to a Poisson distribution, with a mean value $\mu$ assumed to be either known or unknown (in the latter case $\mu$ acts as a nuisance parameter)? As before, the estimate of the cut efficiency is given by the ratio of the measured $k$ and $n$ events, but the determination of the confidence intervals to be attached to the estimate should take into account the occurrence that $n$ fluctuates. It can be anticipated, however, the remarkable result that, with a suitable choice of the acceptance region construction criterion, the confidence intervals actually do not change with respect to the previous case. This is why, for example, in the literature the confidence intervals for the ratio of Poisson means can also be deduced from the Clopper-Pearson limits [12][13][14].

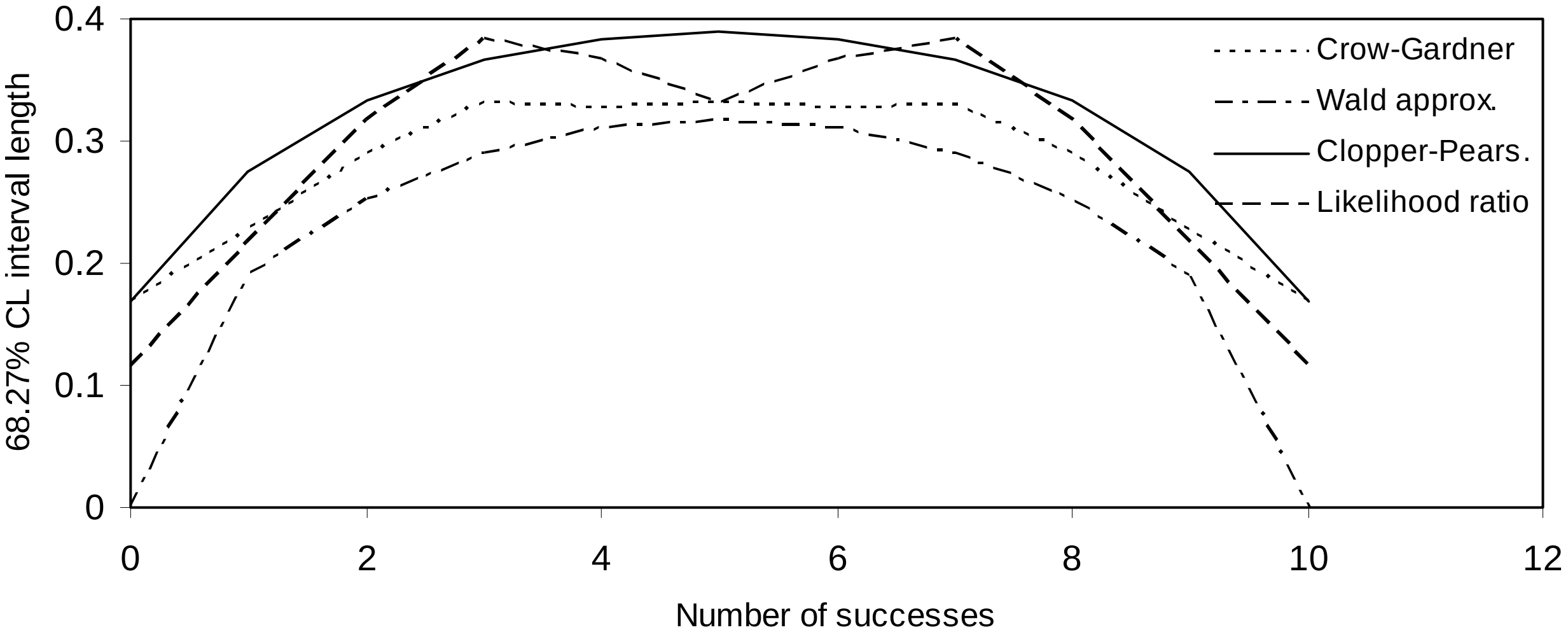

*Fig. 7 – Wald approximation compared with the 68.27% confidence level interval length of the Clopper-Pearson, Crow-Gardner and Likelihood ratio methods for the 10 trials case*

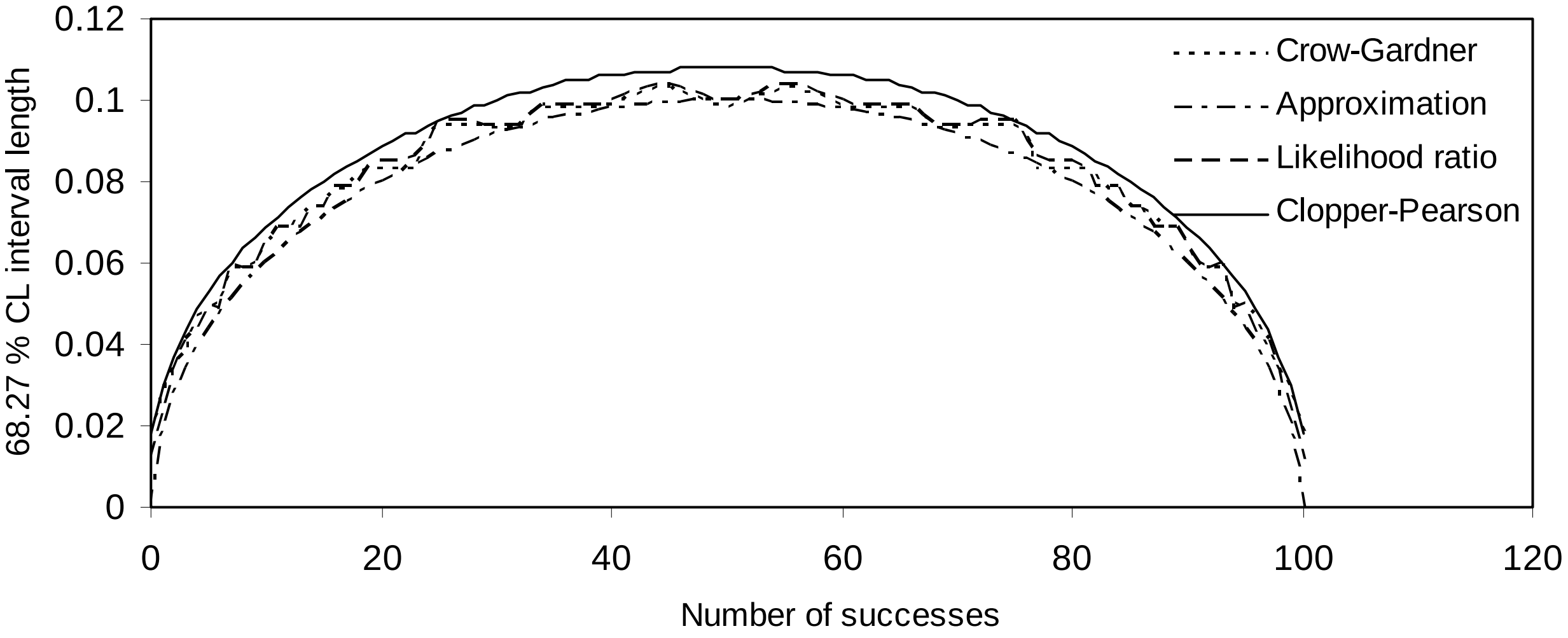

*Fig. 8 – Wald approximation compared with the 68.27% confidence level interval length of the Clopper-Pearson, Crow-Gardner and Likelihood ratio methods for the 100 trials case*

Preliminarily, we note that in the present case the identification of the confidence belt is a tri-dimensional problem. In particular, for every possible value of the success parameter $p$, which assumes values along the $Z$ axis, a bi-dimensional acceptance region in the $X$ and $Y$ plane is constructed by adding points according to some prescription, whose minimum constraint is to ensure the required coverage. Clearly, the choice of the specific prescription determines the resulting confidence intervals.

Specifically, for every point $(n,k)$ in the $X,Y$ plane the probability is computed according to the joint probability expression

$$p(n,k) = p(k/n)p(n) \text{ for } k \le n \tag{4}$$

that is

$$p(n,k)=\binom{n}{k}p^k(1-p)^{n-k}\frac{\mu^n e^{-\mu}}{n!}\ \text{for}\ k\le n \qquad (5).$$

If the acceptance regions are defined imposing separate conditions for each $n$, then the confidence intervals coincide with those obtained for $n$ fixed, irrespective of the true value (known or unknown) of $\mu$.

The demonstration of this result can be easily understood by resorting to the full Neyman construction of the confidence belt; for example, if we want central intervals, with reference to the joint probability formula (5), the acceptance region A($p$) for a given value of the success parameter $p$, evaluated for obtaining 100α% central confidence interval, is obtained by defining the "excluded upper and lower tails" of probability contents β=(1-α)/2 for each value $n$ according to the following formulas

$$\sum_{k=0}^{k_\beta^{down}}\binom{n}{k}p^k(1-p)^{n-k}\frac{\mu^n e^{-\mu}}{n!}\le\beta\frac{\mu^n e^{-\mu}}{n!} \qquad (6)$$

$$\sum_{k=k_\beta^{up}}^{n}\binom{n}{k}p^k(1-p)^{n-k}\frac{\mu^n e^{-\mu}}{n!}\le\beta\frac{\mu^n e^{-\mu}}{n!} \qquad (7)$$

In this way, obviously, while summing the probability content of the excluded regions for all $n$ (i.e. the right terms of the relations (6) and (7)) the factors $\mu^n e^{-\mu}/n!$ sum globally to 1 and thus the upper and lower global tail content is just $\beta$, i.e the design value. On the other hand, (6) and (7) are also trivially simplified removing the equal terms in the right and left members, thus obtaining

$$\sum_{k=0}^{k_\beta^{down}}\binom{n}{k}p^k(1-p)^{n-k}\le\beta \qquad (8)$$

$$\sum_{k=k_\beta^{up}}^{n}\binom{n}{k}p^k(1-p)^{n-k}\le\beta \qquad (9)$$

i.e. the Clopper-Pearson conditions (1) and (2) for $n$ fixed. It must be noted the cancellation of the nuisance parameter, which makes the Clopper-Pearson limit valid irrespective of the true value of $\mu$; the effect of $\mu$ is reflected, instead, into the coverage, which is given by

$$\text{cov}=1-\sum_{k=0}^{k_\beta^{down}}\binom{n}{k}p^k(1-p)^{n-k}\frac{\mu^n e^{-\mu}}{n!}-\sum_{k=k_\beta^{up}}^{n}\binom{n}{k}p^k(1-p)^{n-k}\frac{\mu^n e^{-\mu}}{n!} \qquad (10)$$

So, essentially, this is a very special case in which the nuisance parameter does not influence the confidence limits. The drawback of this result is that, as for the case of $n$ fixed, the obtained limits overcover the true $p$ value much more than the stated confidence level, especially for low $\mu$.

Clearly, the same result is valid for the Crow Gardner and likelihood ratio ordering methods, i.e. the confidence limits are the same as calculated in the previous paragraph, while the coverage depends upon $\mu$ according to the respective formulas corresponding to (10).

To shed more light on this result, we show the nominal 68.27 % coverage for the standard Clopper-Pearson method for the two different μ values 10 and 50 in Fig. 9.

From the figure it can be seen how, in perfect analogy with the case of fixed $n$, the mentioned over-coverage problem, particularly severe for low $\mu$ values, tends anyhow to alleviate as far as $\mu$ increases.

In Fig. 10 we compare the coverage of the three methods for $\mu$ equal to 10. The clear advantage of the Crow-Gardner and likelihood ratio ordering methods is that they substantially mitigate the over-coverage of the standard Clopper-Pearson intervals, though they do not completely eliminate the problem. Also, it should be noted the very similar performances ensured by the two new methodologies.

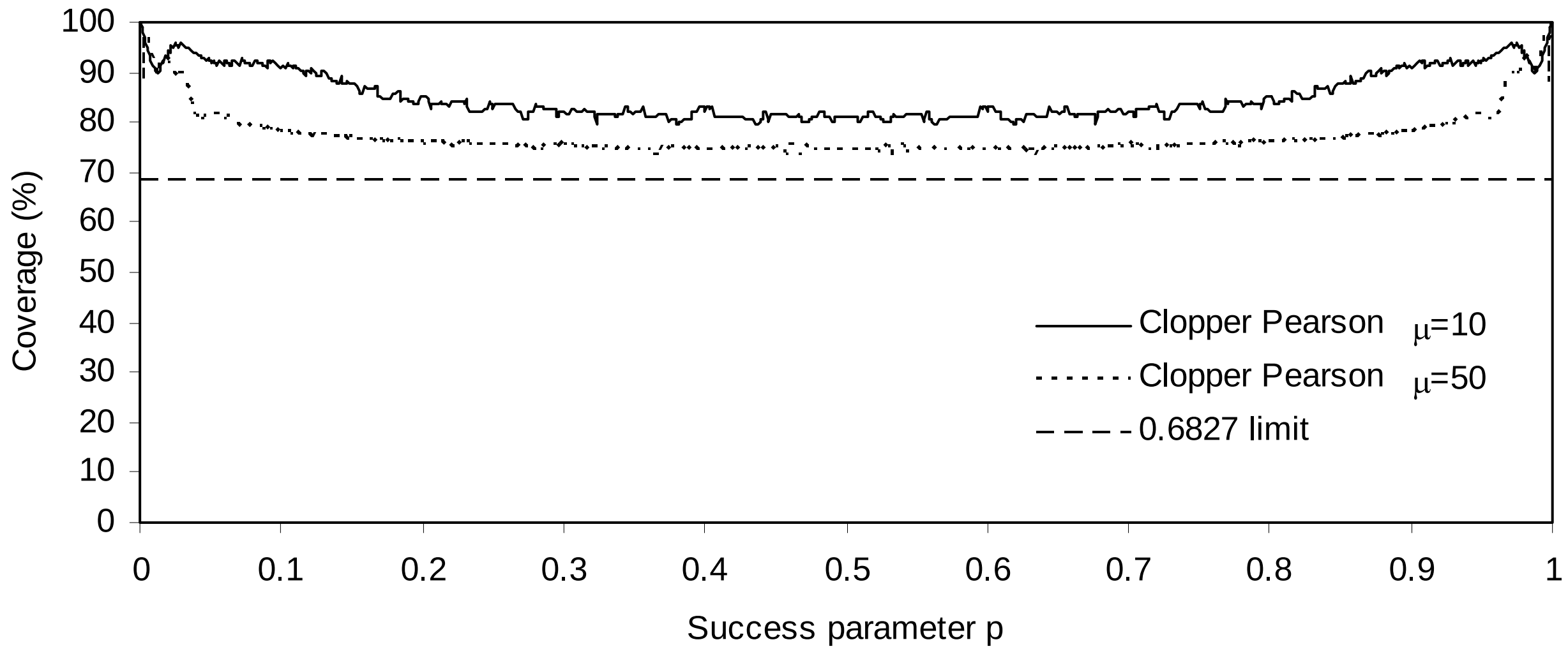

*Fig. 9 – Standard Clopper-Pearson 68.27% confidence level coverage in case of fluctuating number of total events, for the two conditions of $\mu$=10 and $\mu$=50*

In summary, the virtue of the introduction of the limits based on the Crow-Gardner and likelihood ratio orderings is that they ensure a reduced over-coverage with respect to the standard intervals, also when they are used in the case in which the total number of events is not fixed, but subjected to fluctuations.

## 4. Confidence limits and coverage for fluctuating *n* with $\mu$ known

Despite the improvement in term of reduction of the over-coverage obtained through the introduction of the Crow-Gardner and likelihood ratio intervals, the results of the previous paragraph suggest that there is still margin to further reduce significantly the confidence intervals, while keeping the desired constraint on the coverage. In this paragraph we show how this is possible in the case in which the nuisance parameter $\mu$ is known.

To this purpose we should note that the key point in the previous approach was the separate conditions imposed individually on each $n$ while constructing the confidence belt. It is reasonable to guess that a significant reduction in the width of the resulting confidence intervals can be obtained removing this constraint; in other words, for each $p$ the acceptance region is defined with a better optimized strategy in the whole $XY$ plane.

Let's show specifically what happens in such a situation through the adoption of the likelihood ratio ordering. We remind that such a construction means that the acceptance region for each $p$ is obtained via the following criterion

$$\forall p \qquad (n,k) \in A(p) \qquad while \qquad \sum_k \binom{n}{k} p^k (1-p)^{n-k} \frac{\mu^n e^{-\mu}}{n!} \le \alpha \tag{11}$$

being *(n,k)* selected according to the value of ratio

$$r = p(n,k;\mu)/\,lik(n,k;\mu_{best}) \qquad (12)$$

in descending order and as before α is such that 100α% is the desired confidence level. Therefore, the likelihood ratio ordering is performed globally over the plane for each value *p.*

The numerical application of this prescription leads to a tridimensional confidence belt, whose full visualization is somehow difficult. In order to overcome this problem, the confidence belt can be visualized via the set of confidence intervals obtained for varying *k*, for each given value of *n.* For example, in Fig. 11 the length of the 68.27% confidence intervals obtained with this method (dotted line) are reported as computed in the case in which $\mu$ is known to be equal to 10 and for the specific outcome of *n* still equal to 10. It should be emphasized that the entire tridimensional confidence belt could be displayed by plotting all the bidimensional confidence belts for all the possible *n* outcomes from 1 to ∝.

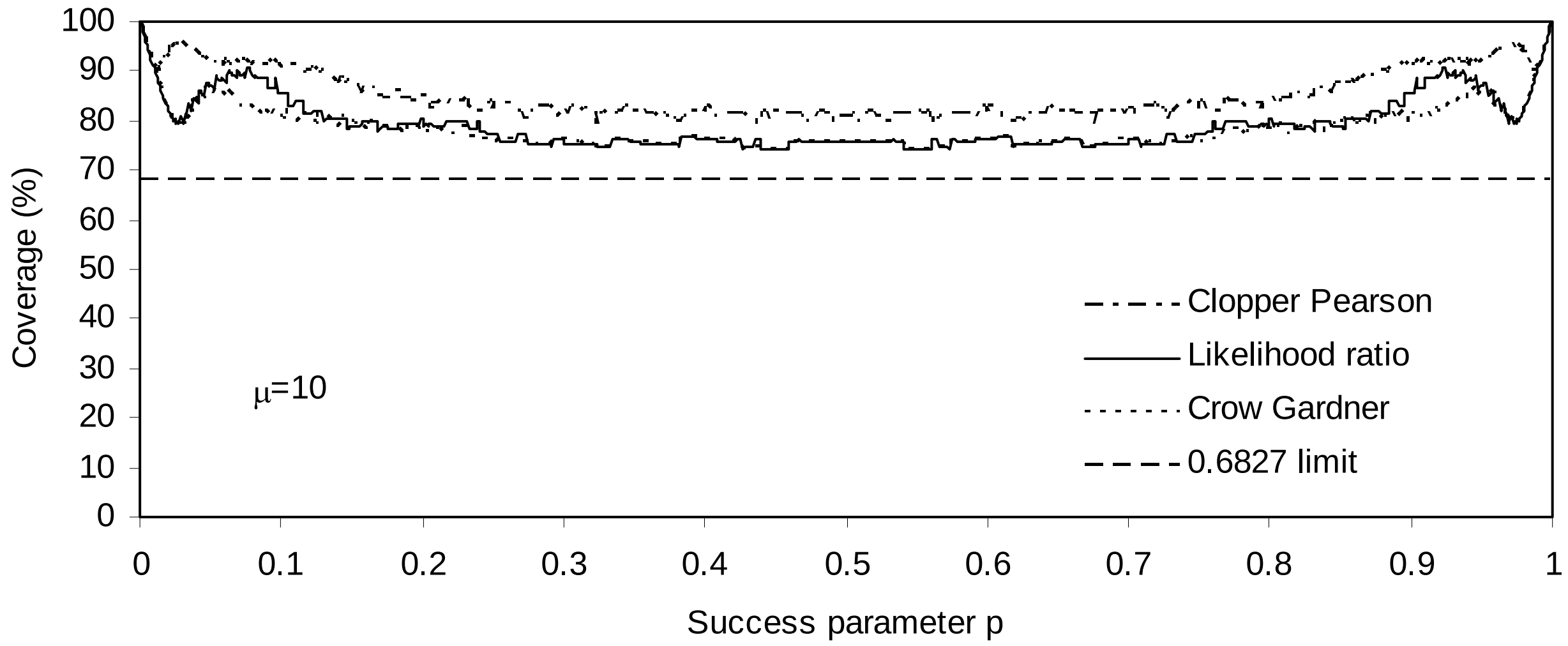

Fig. 10 – Crow-Gardner and Likelihood ratio *68.27% confidence level coverage, in case of fluctuating number of total events, compared with the standard Clopper-Pearson coverage for* $\mu$*=10*

The specific ensemble of confidence interval lengths shown in the figure for the likelihood ratio ordering, as developed in this paragraph, can be conveniently compared with the confidence intervals obtained with the methodologies of the previous paragraphs, for the same 10 total outcomes *n.* In this respect the situation depicted in Fig. 11 is extremely appealing: if one knows the mean background value the resulting confidence intervals are substantially shortened both with respect to the standard Clopper-Pearson case and to the likelihood ratio method applied in the $\mu$ unknown condition, approaching the Wald value.

To gain more insight in the relation between the Wald intervals and the likelihood ratio intervals for $\mu$ known, in Fig. 12 there is the comparison of the relevant confidence belts for the same previous numerical example: while the interval lengths are in general very similar (apart obviously the *k*=0 and *k*=*n* positions), the intervals somehow differ each other, though not too much, since feature non coincident extreme points. Only for *k*=5 there is a practically perfect coincidence.

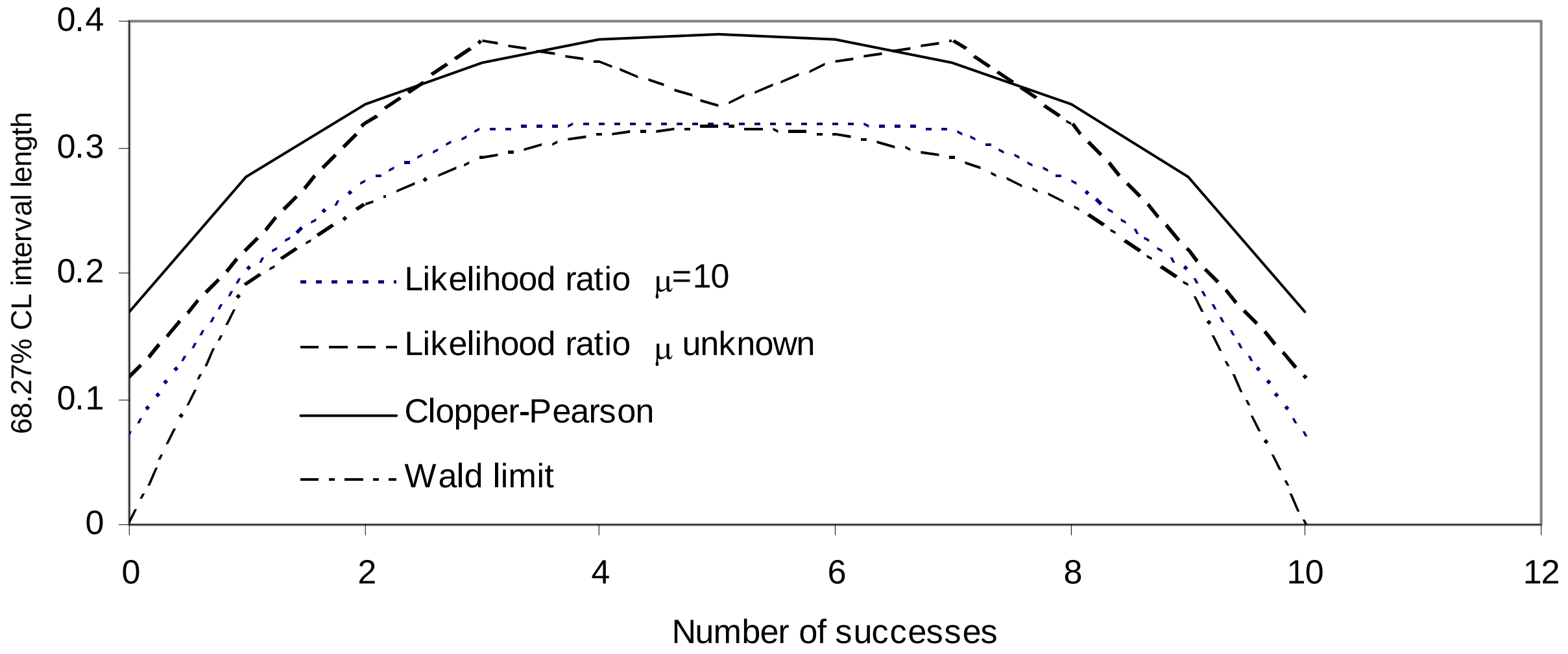

*Fig. 11 – Improved likelihood ratio 68.27% confidence level interval length, for the μ known case, compared with the Wald approximation, the standard Clopper-Pearson and the likelihood ratio for μ unknown*

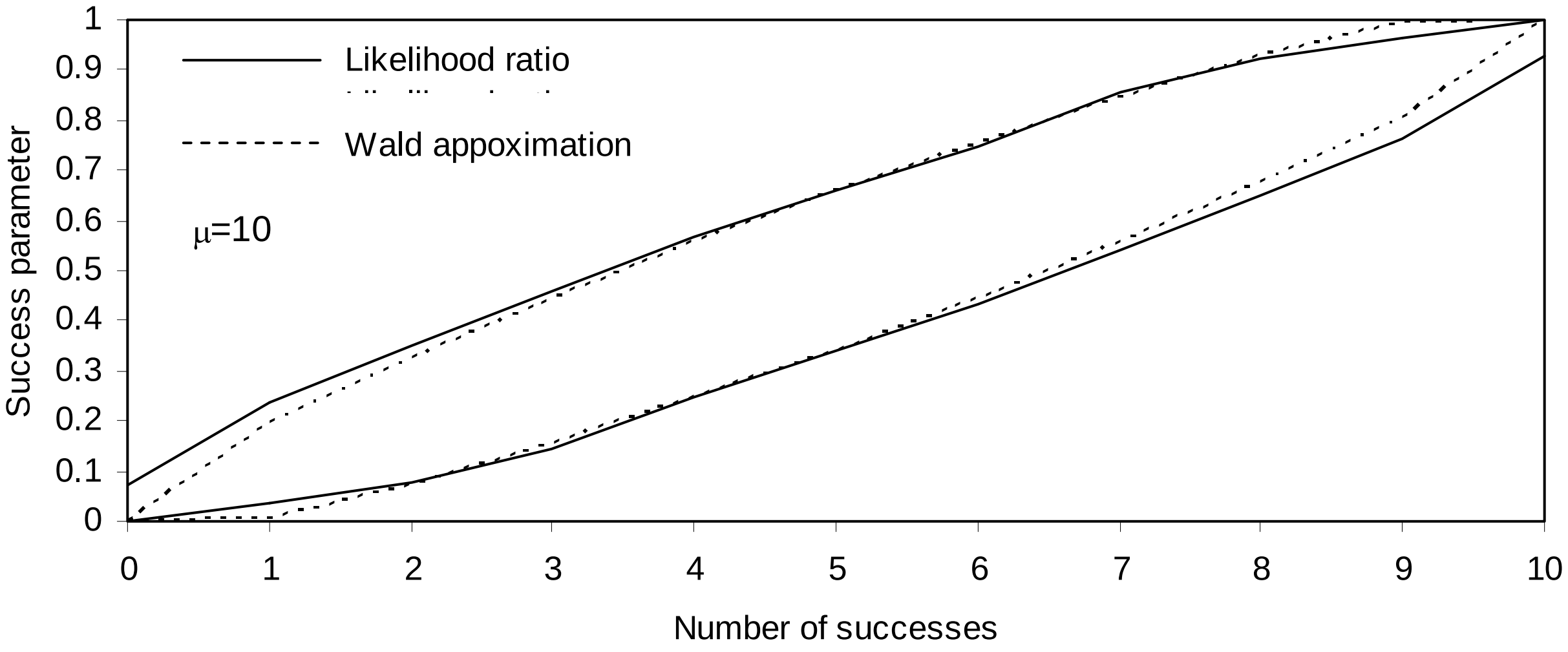

*Fig. 12 – Wald approximation and likelihood ratio with μ known equal to 10: confidence belt comparison*

To shed additional light on the advantage of knowing the background mean value, in Fig 13 a larger subset of the confidence belt is plotted for $\mu$=10, by reporting the confidence interval lengths when the total number of detected counts $n$ is, respectively, 5, 10, 15, 20 and 30. To better interpret the graphs in the figure, it should be underlined that in the tridimensional geometrical construction of the belt it happens that there may be combinations of $k$ and $n$ for which the confidence intervals are empty. In particular, in the specific $\mu$ = 10 and 68.27 % confidence level case addressed here, and for $n$ limited to 30, it stems that for $n>=14$ at the two limiting conditions of $k=0$ or $k= n$ correspond empty intervals. In practice, in order to avoid this unsatisfactory, though legitimate, occurrence when it happens, one can choose to quote larger confidence limits, for example at the 95% confidence level.

Anyhow, it is because of this circumstance that in Fig. 13 the likelihood ratio intervals (dotted lines) are plotted without the two extreme positions of $k=0$ or $k= n$ in three ($n$=15, 20, 30) of the five cases exemplified. In the same figure, instead, the Wald approximation is always plotted with the exclusion of the extreme positions, where it never holds.

With the caveat of the extreme positions in mind, the comparison in Fig. 13 gives the indication that the likelihood ratio ordering performs quite well if the background is known, substantially reducing the larger difference which instead exists between the standard Clopper-Pearson limit and the Wald value.

This interesting result regarding the length of the intervals finds its obvious counterpart also in the coverage: indeed, Fig. 14 shows how the true and design coverage are now remarkably close each other.

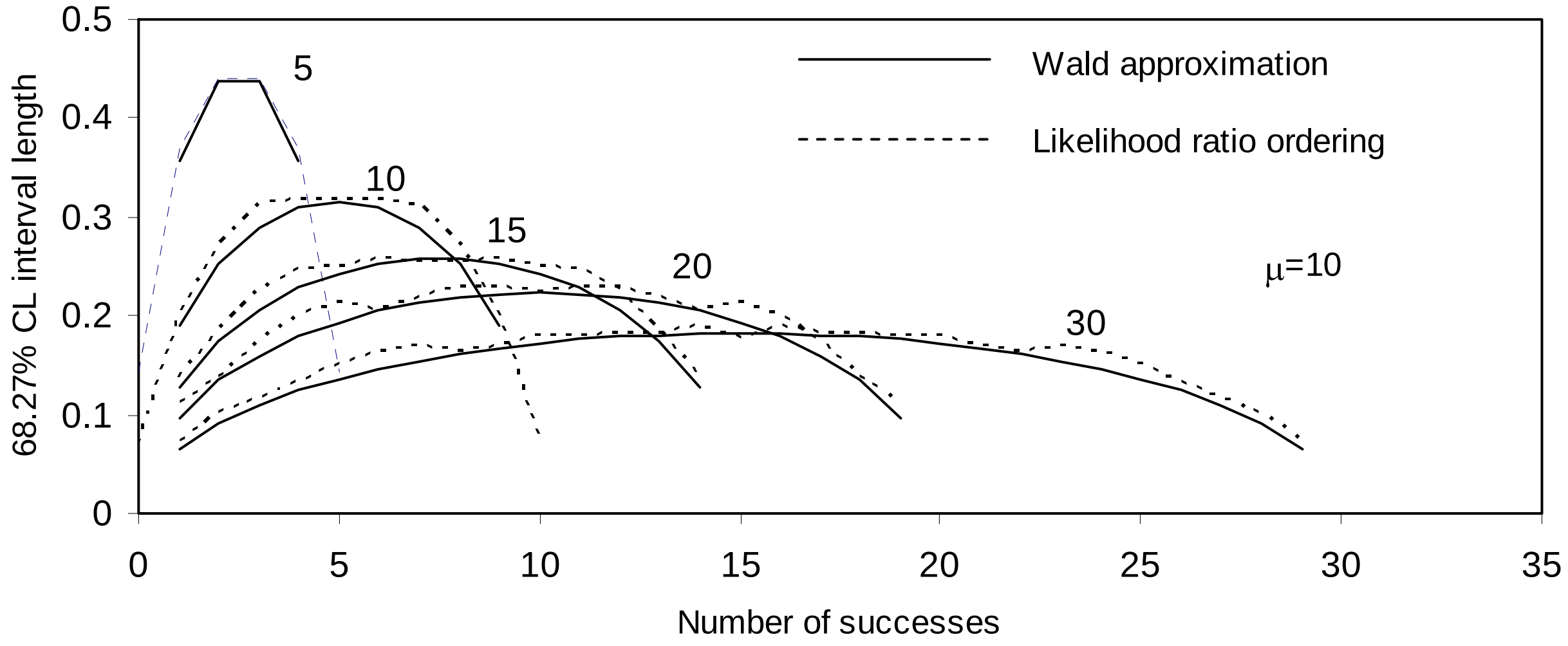

*Fig. 13 – Extended comparison of the confidence interval length for the likelihood ratio method and for the Wald approximation*

The outlined procedure in the case in which the mean background value is known is thus very appealing, producing very short intervals still with the proper coverage. The only, surely tolerable, drawback is that the precise evaluation of the intervals relies on the numerical exploitation of the full frequentist Neyman construction in three dimensions.

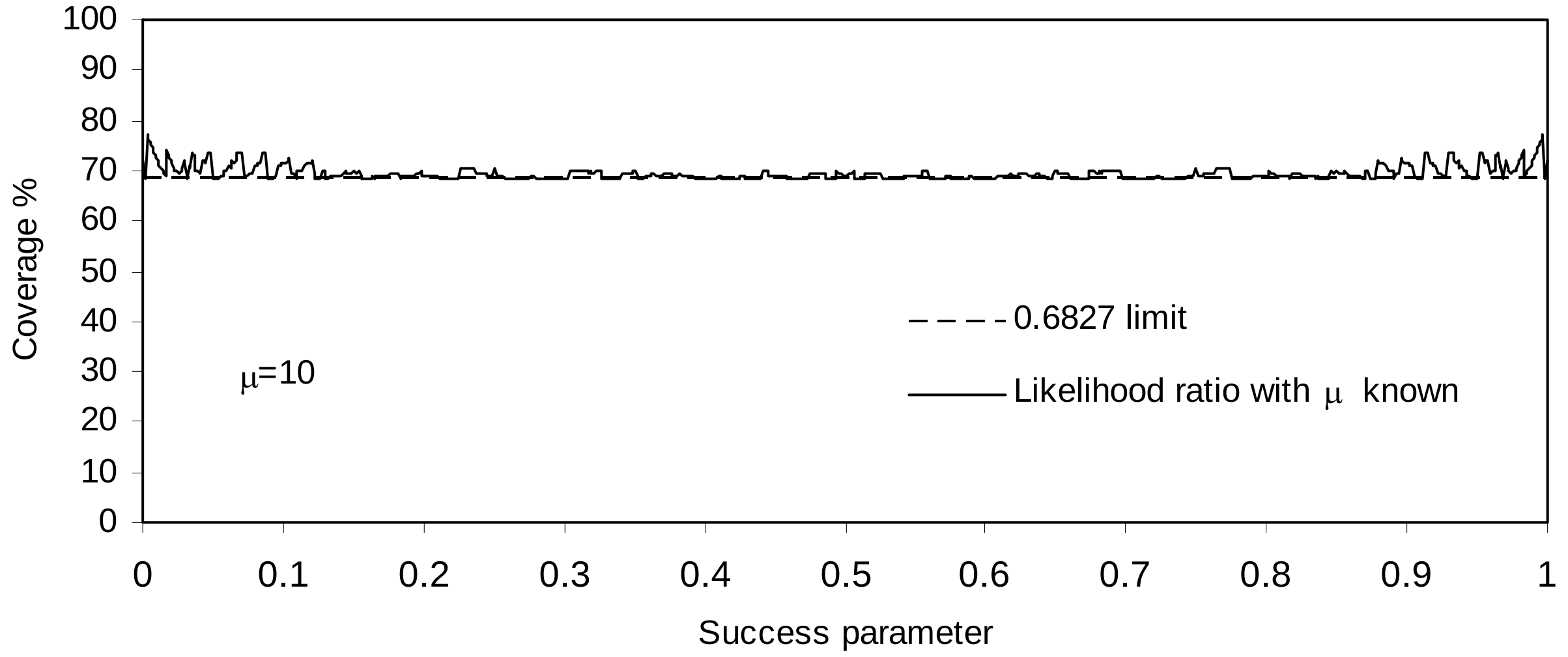

*Fig. 14 – Coverage of the 68.27% confidence intervals produced by the likelihood ratio ordering method, for μ known and equal to 10. It can be appreciated how close the actual and nominal coverage are*

Extensive numerical checks have been carried out for several $\mu$ values in order to test the limits of validity of the previous outcomes. As somehow expected, it has been verified that, even with this methodology, the problem of the over-coverage is not totally eliminated, rather is shifted and confined to very low $\mu$ values. This fact can be appreciated from Fig. 15, displaying the 68.27% coverage for $\mu$ as low as 1, which indeed is generally greater than the nominal coverage.

Furthermore, it has been inferred that when the coverage is close to the nominal value, then the Wald approximation performs reasonably well, while the opposite situation occurs if the coverage is far from the nominal value, as in Fig. 15. As a matter of fact, it results that in the Poisson context the Wald limit provides an acceptable approximation for the length of the confidence intervals of the success parameter $p$ for $\mu \geq 10$, as further discussed in the next paragraph also in terms of implications for the coverage.

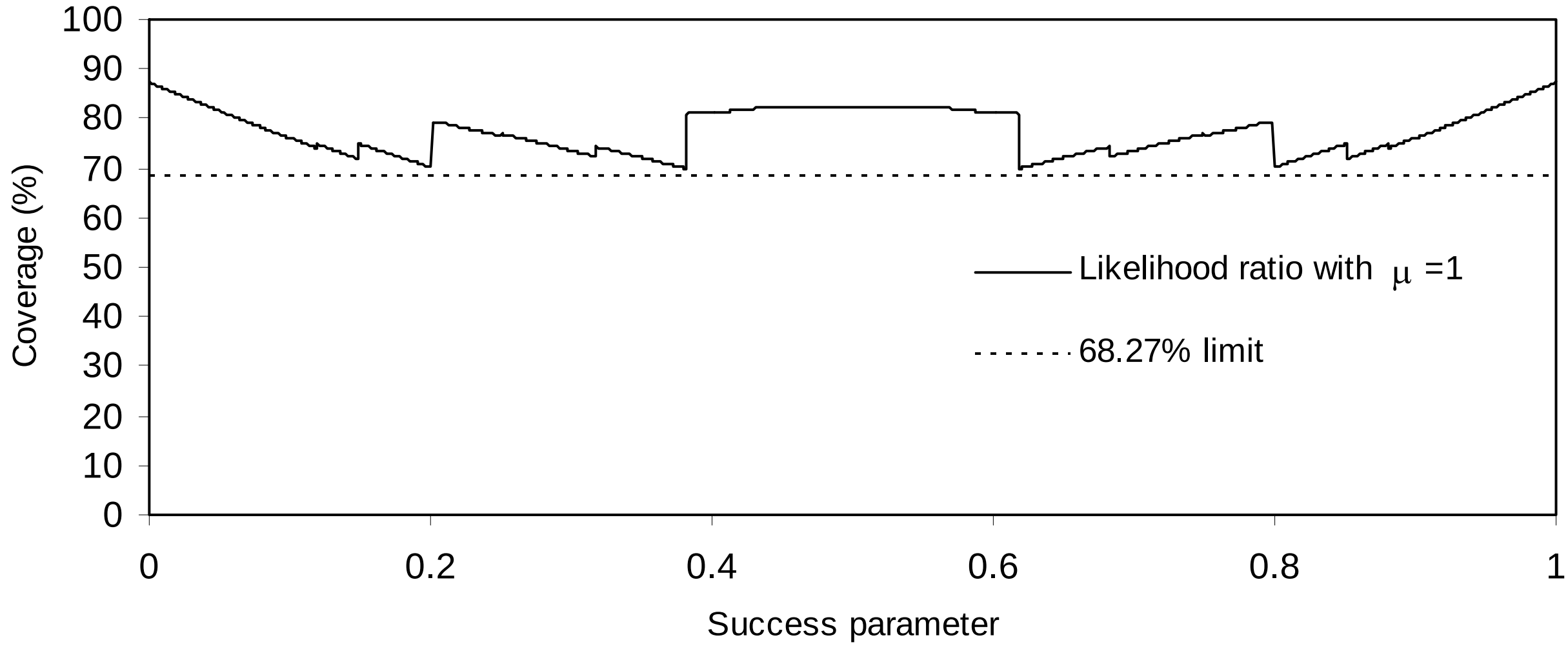

*Fig. 15 – When $\mu$ is very low ($\mu$=1 in this example) a certain amount of over-coverage remains also with the likelihood ratio ordering approach*

At the end of this paragraph the reader may ask why the entire discussion has been focused to the likelihood ratio method, with no mention of the Crow-Gardner approach. The reason is that the latter methodology in the present case would be of no practical usefulness, since it can be easily verified that its application would lead to numerous combinations of $k$ and $n$ outcomes characterized by empty confidence intervals.

## 5. Approximated intervals derived from the likelihood ratio method in the case in which $\mu$ is not known

The derivation in paragraph 4 is based on the premise that the mean value $\mu$ is known, and in this case the problem of the confidence intervals for $p$ is solved in an optimal way. However, since in reality the precise knowledge of $\mu$ is rarely achieved, it may be worth if suboptimal results of practical interest can be obtained from the methodology of the likelihood ratio ordering, even when the mean background value is not known.

In this respect, first of all it should be noted that the Wald approximation is actually independent from $\mu$; consequently a simple and straightforward possibility, whenever the true value of $\mu$ is not accessible, is just to quote the Wald limit. Clearly there is a degree of inaccuracy in this prescription, whose amount can be evaluated by Monte Carlo. Hence for various values of $\mu$ the Wald coverage has been Monte Carlo calculated, with the corresponding results shown in Fig. 16.

The coverage curves in the figure demonstrate that when $\mu$ is low the Wald approximation fails badly. This is not surprising at all, since the Wald large sample test, from which such an approximation is inferred (see next paragraph), is valid just for "large samples". As far as $\mu$ increases, the central portion of the coverage curves tends to reach the nominal 68.27% value, and it does so approaching the theoretical level from the low side. Such a behavior is in agreement with the findings of the previous paragraph, which showed that the Wald approximation is always shorter than the exact intervals.

By inspecting Fig. 16 one concludes that for $\mu \geq 10$ the approximation to the true confidence level in the central portion of the coverage curves can be considered satisfactorily good. On the other hand, whichever is $\mu$, when the success parameter $p$ is close to the extreme values 0 and 1 the coverage drops substantially. Also in this respect the situation improves for the large sample condition, since the interval spanned by the two under-coverage regions shrinks for increasing $\mu$, without, nevertheless, never disappearing completely. This effect is connected to the failure of the Wald approximation for the experimental outcome $k$ equal to 0 or $n$: in fact such occurrences originate empty intervals that, obviously, always fail to cover the true $p$ value.

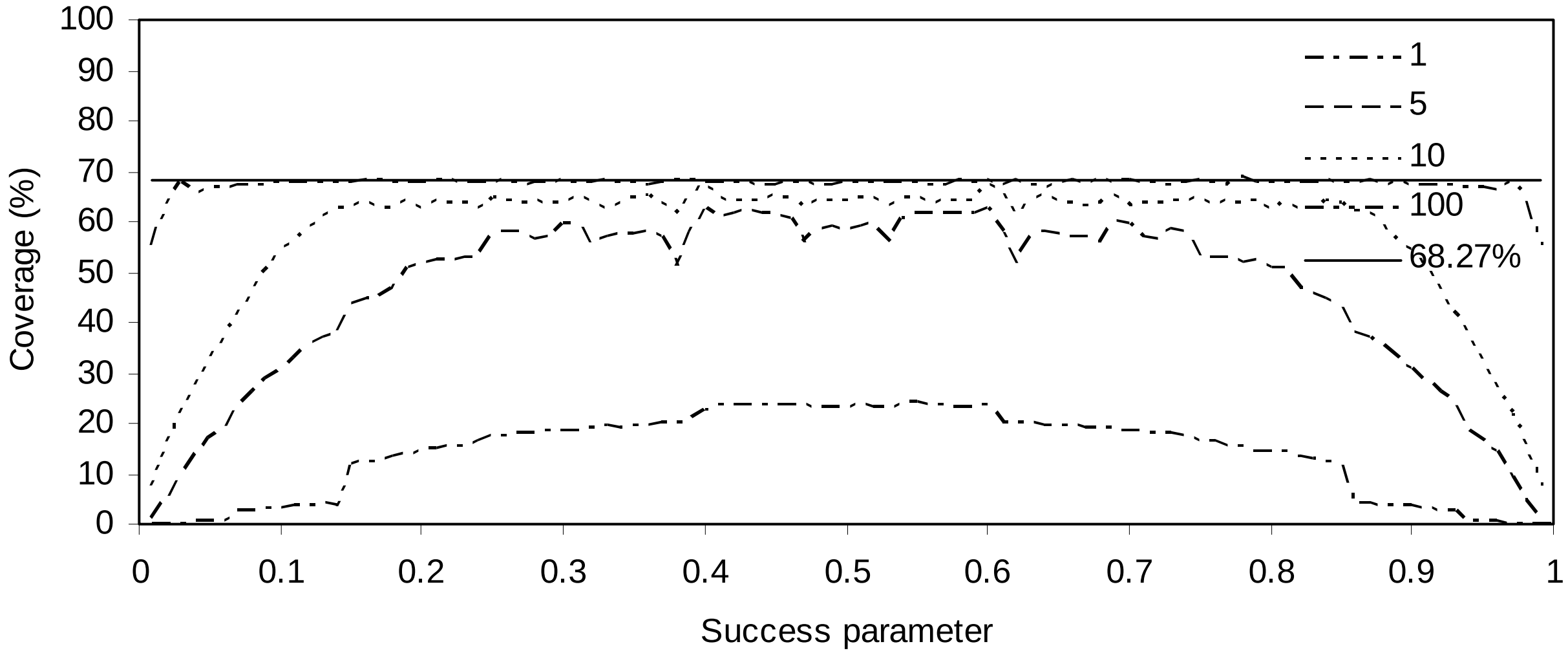

*Fig. 16 – Coverage behavior of the Wald method for various $\mu$ values: if $\mu$ is very low the Wald limit fails badly, on the contrary if $\mu$ $\geq 10$ there is a central flat portion of the coverage curve which ensures a reasonable approximation*

On the basis of the above considerations, a practical recipe to try to exploit the Wald intervals could be the following: if the experimental value $n$ is sufficiently large, at least $\geq 10$, and if the experimentally determined $p=k/n$ estimate of the success parameter falls in the flat portion of the coverage curve evaluated for that $n$, then the Wald approximation can be quoted as a reasonable approximation of the 68.27% confidence interval. Instead, if at least one of the two conditions is not fulfilled, the confidence level to be quoted should be determined via the exact methods of paragraph 3.

Nevertheless, this recipe in practice may be of some usefulness only if $\mu$ is really large, being surely not applicable in the most interesting case of few counts.

Anyhow, inspired by the same procedural approach, one may alternatively want to explore if the very attractive likelihood ratio intervals described in paragraph 4 for $\mu$ known, can be used for the more practical cases in which $\mu$ is not known. This task can be accomplished as follows. If we consider that in an experimental trial we have at our disposal only the values $k$ and $n$, it is conceivable to quote as confidence interval for $p=k/n$ the likelihood ratio interval that is computed for $\mu$=$n$. In other words, $n$ is assumed as best estimate for the unknown $\mu$ value, and on this premise the calculation is carried out as in the paragraph 4 for that estimated value of $\mu$.

Similarly to the Wald approximation, the proposed procedure can be tested by Monte Carlo; Fig. 17 displays the relevant results in term of coverage for $\mu$, respectively, equal to 1, 5, 10 and 100.

At a glance it appears immediately that the situation is much improved with respect to the Wald approximation. For low $n$ there is no more the striking under-coverage shown in Fig. 16, instead, for example for n=1, the methodology features similar (over)coverage properties of the corresponding ideal case of paragraph 4. As far as $n$ increases, not only the coverage approaches the nominal level for most of the values of the success parameter, but also close to the extreme values p=0 and p=1 the severe, not tolerable under-coverage of the Wald approximation is eliminated. Anyhow, clearly these intervals are to be considered as approximated, since the figure shows that there are some values of $p$ characterized by a certain amount of under-coverage: thus they do not strictly fulfill the exact coverage requirement. However, such a mild non ideality is balanced by the advantages ensured by these intervals, which, respect to the exact intervals in paragraph 3 valid for the $\mu$ unknown framework, are of shorter length and with, in average, closer-to-nominal coverage.

In summary, the approximated likelihood ratio intervals introduced in this paragraph, in the case in which the mean background value is unknown, are a suitable alternative to the binomial intervals of paragraph 3, especially in applications for which their slight under-coverage can be tolerated.

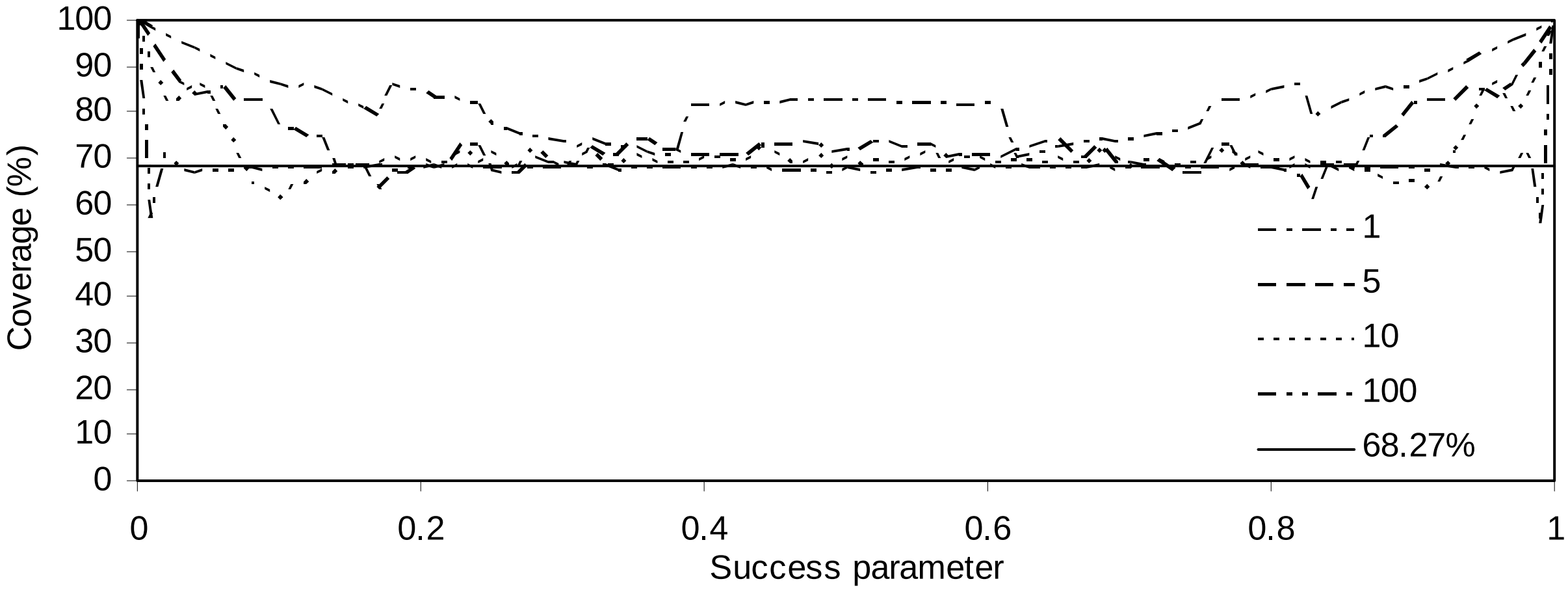

*Fig. 17 – Coverage properties of the likelihood ratio approximated intervals for the same $\mu$ values of the previous figure, which make them a suitable choice in application in which $\mu$ is unknown and a modest under-coverage can be accepted*

## 6. Why the Wald approximation works also in the Poisson case?

The Wald limit is derived in the context of the binomial distribution, but the examples in the previous paragraph show that it provides a satisfactorily approximation also in the Poisson case. It can be interesting to try to demonstrate this property.

We remind that the Wald confidence intervals can be written in the more general formulation as

$$CI_W = \frac{k}{n} \pm \kappa \sqrt{\frac{k}{n^2}\left(1-\frac{k}{n}\right)} \tag{13}$$

where the factor $\kappa$ determines the desired confidence value. $\kappa$ equal to 1 corresponds to 68.27% intervals; if the desired confidence interval is $100\alpha$% and $\beta$ is defined as $\beta=(1-\alpha)/2$, then $\kappa$ is the $1-\beta$ quantile of the standard normal distribution.

The Wald approximation is obtained by inverting the acceptance region of the Wald large sample normal test for a general problem [10] [7] [8] , i.e.

$$\left|(\hat{\theta}-\theta)/\hat{e}(\hat{\theta})\right| \le \kappa \tag{14}$$

where $\theta$ is a generic parameter, $\hat{\theta}$ is the maximum likelihood estimate of $\theta$ and $\hat{e}(\hat{\theta})$ is the estimated standard error of $\hat{\theta}$. In the binomial case the specific relation (13) stems from the condition (14) with the substitutions $\theta=p$, $\hat{\theta}=k/n$ and

$$\hat{e}(\hat{\theta}) = \sqrt{\frac{k}{n^2}\left(1-\frac{k}{n}\right)} \qquad (15).$$

In particular this last relation originates from the variance expression for the variable $k/n$ (remind that in this case $n$ is fixed): simple probabilistic rules tell that the variance is equal to the original binomial variance of the variable $k$ divide by $n^2$ , i.e.

$$\sigma_{k/n}^2 = \frac{1}{n^2}\sigma_{bin}^2 \tag{16}$$

and since

$$\sigma_{bin}^2 = np(1-p) \tag{17}$$

we get

$$\sigma_{k/n}^2 = \frac{1}{n}p(1-p) \tag{18}$$

which coincides with the expression (15) upon substitution of the success parameter $p$ with its experimental estimate $k/n$.

In the case in which $n$ fluctuates it is still true that the equalities $\theta=p$ and $\hat{\theta}=k/n$ hold, while the expression for $\hat{e}(\hat{\theta})$ needs to be re-evaluated. In analogy with the binomial case, this task can be accomplished by preliminarily compute the variance of the random variable $k/n$, taking into account the fluctuation of $n$.

The ingredients for the calculation are the joint probability $p(n,k)$, written as in (5), and the probability $p(k)$, obtained by summing the expression (5) over $n$. The intuitive result that $p(k)$ is a Poisson distribution with parameter $\mu'=p\mu$, can be also rigorously demonstrated (see for example [15] and [16]).

The covariance of the two variables $n$ and $k$ can be written as

$$\mathrm{cov}(n,k) = E(nk) - E(n)E(k) \tag{19}$$

where the symbol $E$ denotes the expectation value of the variable within brackets, being obviously $E(n)=\mu$ and $E(k)=p\mu$ .

$E(nk)$ can be evaluated resorting to the characteristic function $f_c(s_1,s_2)$ of the joint probability $p(n,k)$ (5), i.e.

$$f_c(s_1,s_2) = \sum_{n=k}^{\infty}\sum_{k=0}^{\infty} s_1^n s_2^k p(n,k) \tag{20}$$

which after some manipulations becomes

$$f_c(s_1,s_2) = e^{-\mu + s_1\mu - ps_1\mu + s_1 s_2 p\mu} \tag{21}$$

Because of the properties of the characteristic function, $E(nk)$ is equal to the first mixed derivative of relation (21) evaluated for $s_1=s_2=1$

$$E(nk) = \frac{\partial^2 f_c(s_1,s_2)}{\partial s_1 \partial s_2} \quad \text{for } s_1=s_2=1 \tag{22}$$

By carrying out the derivative we get

$$E(nk) = p\mu + p\mu^2 \tag{23}$$

and by inserting (23) in the relation (19) we obtain

$$\mathrm{cov}(n,k) = p\mu + p\mu^2 - p\mu^2 = p\mu \tag{24}$$

Having obtained this result, we can now write the error propagation formula for $p=k/n$, when both $k$ and $n$ fluctuates. We have

$$\sigma_p^2 = \left(\frac{\partial p}{\partial n}\right)^2 \sigma_n^2 + \left(\frac{\partial p}{\partial k}\right)^2 \sigma_k^2 + 2\frac{\partial p}{\partial n}\cdot\frac{\partial p}{\partial k}\cdot \mathrm{cov}(n,k) \tag{25}$$

being the derivatives computed for $n=\mu$ and $k=p\mu$.

By replacing $\frac{\partial p}{\partial n} = -\frac{p}{\mu}$, $\frac{\partial p}{\partial k} = \frac{1}{\mu}$, $\sigma_n^2 = \mu$, $\sigma_k^2 = p\mu$ and $\mathrm{cov}(n,k) = p\mu$ we get

$$\sigma_p^2 = \frac{p}{\mu}(1-p) \tag{26}$$

and by substituting $\mu$ with $n$, i.e. the actual measured number, and $p$ with its experimental estimate $k/n$, we obtain for the estimated variance

$$\hat{\sigma}_p^2 = \frac{k}{n^2}\left(1-\frac{k}{n}\right) \tag{27}$$

i.e. the same expression valid for the binomial case. It is because of this equality that the same Wald approximation applicable in the pure binomial case maintains its validity in the occurrence that $n$ is Poisson distributed. From a pure mathematical point of view, the reason why it works also when $n$ fluctuates can be traced back to the effect of the covariance term in the error propagation formula.

## 7. Connection with the confidence intervals for the ratio of Poisson means

In paragraph 3 the connection of the methodology described in this paper for the confidence intervals of cut efficiencies when the sample population fluctuates, with the classical problem of setting limits on the ratio of two Poisson means has been already mentioned. Here this point is further elucidated.

Given two Poisson ensembles characterized, respectively, by the mean values $\mu_1$ and $\mu_2$, upon sampling $k_1$ instances from the first ensemble and $k_2$ instances from the second ensemble it is desired to set a confidence interval for the ratio $\lambda=\mu_1/\mu_2$. The problem can be rephrased by imagine to merge the two ensembles in a unique ensemble, thus characterized by mean value $\mu=\mu_1+\mu_2$, and therefore asking to set a confidence interval on the proportion $p$ between the events belonging to the sub-ensemble 1 and the total events in the global ensemble. In this way the problem is made equal to the determination of the limits of a binomial proportion $p$ when the total population fluctuates, as addressed in paragraphs 3 and 4; specifically, upon identifying $k_1 \rightarrow k$ and $k_1+k_2 \rightarrow n$ according to the previous designations, then the joint probability $p(k,n)$ can be written as expression (5), being $p$ given by $\mu_1/(\mu_1+\mu_2)$ (a formal demonstration of this quite intuitive result can be found in [13], [17] and [18]). Therefore, the desired confidence interval for $p$ is evaluated as per one of the approaches explained in the above paragraphs, and denoting its upper and lower extremes with $p_{up}$ and $p_{down}$, consequently the confidence interval on $\lambda$ has bounds given by $p_{up}/(1-p_{up})$ and $p_{down}/(1-p_{down})$.

This is exactly the procedure suggested by James and Roos in [5]. In the general case in which $\mu=\mu_1+\mu_2$ is not known, the demonstration in paragraph 3 justifies the evaluation of the limits on $\lambda$ starting from the intervals inferred in the framework of the pure binomial problem (in [5], indeed, the procedure is based on the Clopper-Pearson intervals). Instead, if $\mu=\mu_1+\mu_2$ is known, then shorter intervals on $\lambda$ can be assessed via the limits on $p$ inferred through the three-dimensional frequentist approach of paragraph 4. The very attractive approximated intervals of paragraph 5 can be adopted, as well, in the practical cases in which $\mu=\mu_1+\mu_2$ is not known.

## 8. Conclusions

Motivated by the cut efficiency problem stated at the beginning, this paper generalizes on it, leading to the introduction of novel confidence intervals for the proportion of a binomial distribution and for the ratio of Poisson numbers, with very attractive features regarding both coverage and length. Indeed, all the described exact intervals alleviate the well known over-coverage features of the classical Clopper-Pearson solution and are generally of shorter length, still maintaining the desired property to ensure no under-coverage for any combination of the true parameter $p$ and of the experimental outcomes. In particular, the binomial Crow-Gardner and likelihood ratio intervals developed in paragraph 2 and 3 can be usefully applied in the pure binomial framework when $n$ is fixed, as well as in the Poisson case for fluctuating $n$, with beneficial effects on the coverage properties in both instances.

When the sample population $n$ fluctuates with known mean value $\mu$, the combination of the straightforward classical Neyman prescription with the likelihood ratio ordering procedure originates extremely attractive intervals, that ensure virtually the nominal coverage. Furthermore, such intervals compare very favorably with the widely used Wald approximation, classically derived in the framework of the binomial problem. The pedagogical virtue of these intervals must be underlined, too, since they stem from the canonical application of the full tridimensional frequentist construction of confidence intervals, via the concept of likelihood ratio ordering. Moreover, by releasing the strict coverage requirements, the likelihood ratio ordering produces approximated intervals, for the $\mu$ not known case, that can be of significant practical interest in applications in which a modest under-coverage can be tolerated.

By virtue of their superior properties, the confidence intervals described in this paper should be taken in due consideration as complement and alternative possibilities with respect to the standard Clopper-Pearson intervals reported by the PDG. The complication associated to their determination via the Neyman construction is surely more than counterbalanced by the advantages that they guarantee in term of more precise coverage and of shorter length.